\documentclass[a4paper,12pt]{article}

\usepackage[T2A]{fontenc}
\usepackage{amsmath,amssymb}
\usepackage{cite}
\usepackage[unicode,linktocpage=true,plainpages=false,pdfpagelabels=false]{hyperref}

\newcommand{\dd}{\partial}
\newcommand{\de}{\delta}
\newcommand{\m}{\mu}
\newcommand{\n}{\nu}
\newcommand{\ls}{\left(}
\newcommand{\rs}{\right)}
\newcommand{\La}{\Lambda}
\newcommand{\ka}{\varkappa}
\newcommand{\ga}{\gamma}
\newcommand{\ff}{\varphi}
\newcommand{\ta}{\tau}
\newcommand{\al}{\alpha}
\newcommand{\be}{\beta}
\newcommand{\Ga}{\Gamma}
\newcommand{\ps}{\psi}
\newcommand{\np}{\emptyset}

\newcommand{\disn}[2]{$$\displaylines{\refstepcounter{equation}%
		\label{#1}\hskip 1em minus 1em #2\hfilneg}$$}
\newcommand{\nom}{\hfil\hskip 1em minus 1em (\theequation)}
\newcommand{\no}{\hfil \hskip 1em minus 1em\phantom{(\theequation)}%
	\hfilneg\cr\hfilneg\hskip 1em minus 1em\hfil}
\newcommand{\ns}{\hfill\cr\hfill}

\begin{document}

\title{Dark matter from non-relativistic embedding gravity}

\author{S.~A.~Paston\thanks{E-mail: pastonsergey@gmail.com}\\
{\it Saint Petersburg State University, Saint Petersburg, Russia}
}
\date{\vskip 15mm}
\maketitle

\begin{abstract}
We study the possibility to explain the mystery of the dark matter through the transition from General Relativity to embedding gravity. This modification of gravity, which was proposed by Regge and Teitelboim, is based on a simple string-inspired geometrical principle: our spacetime is considered here as a 4-dimensional surface in a flat bulk.
We show that among the solutions of embedding gravity, there is a class of solutions equivalent to solutions of GR with an additional contribution of non-relativistic embedding matter, which can serve as cold dark matter.
We prove the stability of such type of solutions and obtain an explicit form of the equations of motion of embedding matter in the non-relativistic limit.
According to them, embedding matter turns out to have a certain self-interaction, which could be useful in the context of solving the core-cusp problem that appears in the $\La$CDM model.
\end{abstract}

\newpage

\section{Introduction}\label{vved}
The mystery of the dark matter (DM) is one of the most intriguing problems of contemporary physics. The hypothesis of the DM existence allows us to explain multiple observational discrepancies at different scales: from galactic to cosmological ones, see, e.g., \cite{1611.09846}.  Essentially, it explains the deviation from the expected motion of stars ("rotation curves") at the galactic scale. At the larger scale, it explains the results of observations of the gravitational lensing and the baryon acoustic oscillations. Finally, at the cosmological scale, it solves the problem of structure formation and takes part (along with dark energy) in solving the problem of the deficit of the total Universe mass in comparison with the value corresponding to the critical density.

In general, all present observations (among which the observation of cosmic microwave background anisotropy plays the most significant role) can be well described by the $\La$-Cold Dark Matter ($\La$CDM) model
\cite{gorbrub1}, which today serves as a standard model of the cosmology. In its framework, DM can be considered as non-relativistic dust matter, which generates the same gravitational field that ordinary matter, but the non-gravitational interaction of DM and ordinary matter either absent or negligibly small. For the basic DM properties that follow from observations, see \cite{2005.03520}.

The significant number of situations in which the introduction of DM proves itself helpful allows us to suppose that it very likely exists, even though all attempts to detect it have been fruitless \cite{1509.08767,1604.00014}. Probably the most recently popular ideas of the description of the DM are assumptions that DM is Weakly Interacting Massive Particles (WIMPs) \cite{1703.07364} or fuzzy DM \cite{astro-ph/0003365}. The failure of the attempts of direct DM detection in these models is associated with the weakness of interaction between DM and ordinary matter. Among the others, there are models in which DM is self-interacting (SIDM) \cite{1705.02358}, which provides a possible way to solve the problem of excess density in the galactic cores (the core-cusp problem).

However, the fact that no one has succeeded in detecting any interaction of DM besides gravitational suggests that it is not a real matter but rather a specific effect of the description of the gravitational interaction, i.e., within the framework of the fundamental theory, DM does not \textit{really} exists. It was done for the first time, probably, in the framework of the MOND paradigm \cite{mond},
for the review of attempts to explain observations in the MOND framework see, e.g., \cite{2007.15539} and the references therein.
In this approach, it is possible to successfully eliminate the abovementioned discrepancies at the galactic (or not much larger) scale, but MOND does not work so well at the cosmological scale \cite{1910.04368}. More promising seem to be modified gravity theories, which possess additional degrees of freedom in relation to GR, which correspond to DM when the theory described in terms of GR, e.g., $f(R)$ gravity, scalar-tensor gravity, and others,
see the review of modifications in \cite{1108.6266}, see also \cite{1708.00603} and \cite{2007.15539}).
It should be stressed that after the transition to the modified theory, DM \textit{no longer exists} as a separate entity. Only if we rewrite the theory in the form of GR, DM appears as a source of additional contribution in the Einstein equations.

A new model of such type has drawn much attention in recent years \cite{mukhanov}, namely the mimetic gravity, which appears as a result of the change of variables in the GR that isolates the conformal mode of the metric.
Since this change of variables contains differentiation and is non-invertible \cite{arXiv1311.3111,arXiv1407.0825},
the modified theory resulting from the change contains, in addition to solutions of the Einstein equations, additional solutions, i.e., the extension of dynamics occurs as a result of differential transformations of field variables \cite{statja60}.
The action of the mimetic gravity can be rewritten as the action of GR with some additional term \cite{Golovnev201439}, so it can be seen that in the mimetic gravity, DM is effectively parametrized by two scalar fields, being a pressureless ideal fluid with potential motion.
There are also many other ways to write an action for mimetic gravity,
see, e.g., \cite{arXiv1311.3111,arXiv1512.09118}.
The restriction of the potentiality of motion can be lifted \cite{statja48}, the introduction of a pressure is also possible \cite{lim1003.5751},
but nevertheless the mimetic gravity needs to be significantly augmented
to successfully resolve the issues mentioned above,
see, e.g., \cite{1601.00102} and also review \cite{mimetic-review17}.

The mimetic gravity is not the only variant of modified gravity, which appears as a result of the change of variables in the GR. Another theory of such kind is the so-called embedding theory (or "embedding gravity"), which was proposed by Regge and Teitelboim \cite{regge} and in the subsequent years was discussed in \cite{deser} and many other works (see, e.g., references in  \cite{statja26}). It shares some features with mimetic gravity but has deep geometric origins in contrast with it. In this string-inspired approach, our spacetime is considered as a surface in flat ambient space (bulk), and its metric becomes induced, i.e., the change of variables mentioned above has the form
\disn{r1}{
	g_{\m\n}=(\dd_\m y^a)(\dd_\n y^b)\,\eta_{ab},
	\nom}
where $y^a(x)$ is an embedding function which described the shape of the surface, and $\eta_{ab}$ is a flat bulk metric
(here and hereafter $\m,\n,\ldots=0,1,2,3$; $a,b,\ldots=\np,1,\ldots,9$).
In this framework, one can try to explain the effects of DM in the assumption of strict homogeneity and isotropy of the universe \cite{davids01,statja26}. In order to do the same at a smaller scale, it is useful to reformulate the theory as GR with an additional DM contribution to the action \cite{statja51}, so the analysis of the properties of this fictitious \emph{embedding matter} becomes more simple. The number of variables parameterizing the embedding matter turns out to be quite large, and its properties are rich enough to expect successful modeling of observed DM without additional extensions (which are required in the mimetic gravity approach). The fact that the embedding theory is based on a simple geometric idea without the introduction of artificial extensions and arbitrary constants in the action, together with the presence of the flat ambient spacetime in the theory, which could be of use in attempts to quantize it (see, e.g., \cite{pavsic85,davkar}) makes this modification of gravity, in our opinion, an especially promising candidate for the explanation of DM mystery.

In this work, we study the class of solutions of embedding gravity correspondent to the situation, in which the conserved currents describing embedding matter are non-relativistic vectors in the bulk.
It can be shown \cite{statja68} that for such solutions, the motion of embedding matter in our four-dimensional spacetime is non-relativistic in a certain coordinate system, in which the metric is close to the flat one.

In section~\ref{emb} we write (following \cite{pavsic85}) the equations of motion for embedding gravity as Einstein equations (with additional contribution from embedding matter) together with equations of motion of this matter. However, our independent variable is not an embedding function, but a non-square vielbein, as was proposed in the work \cite{faddeev}, instead of embedding function. This makes the approach of the present work different from the approach in the paper \cite{statja68}. In the section~\ref{sv-din} we split the set of obtained equations into dynamical ones and the constraints imposed at the initial moment.
Such a division simplifies the subsequent analysis of the equations of motion.
In section~\ref{nonrel} we study the appearing system of constraints and dynamical equations in the abovementioned non-relativistic limit, supposing that the gravitational field is weak.
As a result,
we prove the stability of the considered
class of solutions in such a sense that if the initial values correspond to this class, the solution remains in it during the evolution.
It should be stressed that this result is nontrivial, since in this theory there are no parameters characterizing the coupling of embedding matter that could be artificially made small.

In that way,
the non-relativistic regime of embedding matter motion turns out to be stable, and we find the corresponding equations of motion, which lead to the presence of certain self-interaction of embedding matter.
If this self-interaction is not too strong (it depends on the choice of initial values for the epoch of its non-relativistic motion), such behavior is in good agreement with known properties of the DM at all scales,
and the very existence of this self-interaction might help to solve the abovementioned core-cusp problem.
The further study of obtained non-relativistic equations of motion
(numerical simulations might be required)
can help to solve the question of the behavior of embedding matter in different regimes.
The comparison of such results with observations should show the effectiveness of the present approach to the description of dark matter in solving the core-cusp problem and other
inconsistencies \cite{2007.15539} related to the results of
N-body simulations in the $\La$CDM model on the galactic scale,
such as the missing satellites and too-big-to-fail problems.

\section{Embedding gravity}\label{emb}
In the embedding gravity the role of the independent variable describing the gravitational field is played by the embedding function
$y^a(x)$. The action is Einstein-Hilbert one with arbitrary matter, in which the metric in the form  \eqref{r1} is then substituted. The variation of this action w.r.t. $y^a$ leads to the Regge-Teitelboim equations \cite{regge}:
\disn{r2}{
	D_\m\Bigl(( G^{\m\n}-\ka\, T^{\m\n}) \dd_\n y^a\Bigr)=0,
	\nom}
where $D_\m$ is a covariant derivative, $G^{\m\n}$ is the Einstein tensor and $T^{\m\n}$ is the EMT of an ordinary matter. These equations can be rewritten \cite{pavsic85} as the following system of equations
\disn{r3}{
	G^{\m\n}=\ka \ls T^{\m\n}+\ta^{\m\n}\rs,\qquad
	D_\m\Bigl(\ta^{\m\n}\dd_\n y^a\Bigr)=0,
	\nom}
which can be interpreted as the Einstein equations with the additional contribution $\ta^{\m\n}$ to the EMT from the embedding matter, along with the equation of motion of the latter. This equation looks like a conservation law for the set of currents
\disn{s6}{
	D_\m j^\m_a=0,\qquad
	j^\m_a=\ta^{\m\n}\dd_\n y_a.
	\nom}
If the embedding gravity is reformulated as GR with some additional matter (in this approach, the metric $g_{\m\n}$ is treated as an independent variable, and the condition \eqref{r1} appears as one of the Euler-Lagrange equations), these currents, alongside with embedding function, can be chosen as independent variables describing embedding matter (the corresponding action was constructed in \cite{statja51}). Alternatively, the symmetric quantity $\ta^{\m\n}$ can be chosen instead of $j^\m_a$ \cite{statja48}.

The set of currents $j^\m_a$ can be treated as four vectors (counted by index $\m$) in ten-dimensional bulk with the signature $(+-\ldots-)$. The main aim of this work is to study the equations of motion of embedding matter in such limit that all these vectors are \emph{non-relativistic in the bulk}, i.e.
\disn{s7}{
	j^\m_a=\de^\np_a j^\m+\de j^\m_a,\qquad
	\de j^\m_a\to0.
	\nom}

It turns out to be helpful in the analysis of the equations to use non-square vielbein  $e^a_\m\equiv\dd_\m y^a$ as an independent variable instead of $y^a$, imposing the integrability condition on it
\disn{s8}{
	D_\m e_\n^a=D_\n e_\m^a
	\nom}
(here, the covariant derivative $D_\m$ was used for the convenience, but it is equivalent to the use of the ordinary derivative $\dd_\m$). Note that in the context of embedding theory, the usage of non-square vielbein as an independent variable was proposed in the work \cite{faddeev}.

As a result, the equations of motion of embedding matter can be written as
\disn{s9.1}{
	e_\m^a e_\n^b\,\eta_{ab}=g_{\m\n},\qquad
	D_\m e_\n^a=D_\n e_\m^a,
\nom}\vskip -2em
\disn{s9.2}{
	D_\m\Bigl(\ta^{\m\n}e_\n^a\Bigr)=0,
\nom}
in which the variables $e_\m^a$ and $\ta^{\m\n}$ (one can use $j^\m_a$ as an independent variable instead of $\ta^{\m\n}$, but in this case the form of the equations is slightly different \cite{statja51}) describe embedding matter, and the metric $g_{\m\n}$ can be found as a solution of the Einstein equation that is accounted separately (see \eqref{r3}).

\section{Constraints and dynamical equations}\label{sv-din}
Let us extract from the set of equations \eqref{s9.1},\eqref{s9.2} the ones that we will call constraints: these are the equations that do not contain time derivatives $\dd_0\equiv\partial/\partial x^0$ of embedding matter variables $e_\m^a$, $\ta^{\m\n}$ (although we assume that they can contain $\dd_0 g_{\m\n}$). The remaining equations we will call dynamical ones.

To start with, we note that equations \eqref{s9.1} lead to a relation that is known in the embedding theory (see, e.g., \cite{statja18}):
\disn{s10}{
	e_{\be a} D_\m e_\n^a=0.
	\nom}
It allows to rewrite equation \eqref{s9.2} as two equations
\disn{s11}{
	D_\m \ta^{\m\n}=0,\qquad
	\ta^{\m\n}D_\m e_\n^a=0.
	\nom}
As a result, we can rewrite \eqref{s9.1},\eqref{s9.2} as a combination of constraints
\disn{s13}{
	e_\m^a e_{\n a}=g_{\m\n},\qquad
	D_i e_k^a=D_k e_i^a
	\nom}
and dynamical equations
\disn{s14}{
	D_0 e_k^a=D_k e_0^a,\qquad
	D_0 \ta^{0\n}=-D_i \ta^{i\n},\qquad
	D_0 e_0^a=-\frac{1}{\ta^{00}}\ls 2\ta^{0i}D_i e_0^a+\ta^{ik}D_i e_k^a\rs,
	\nom}
where $i,k,\ldots=1,2,3$.

Now it is possible to apply the procedure closely related to the one that is used in the Hamiltonian description of constrained systems: when one takes the time derivative of the appearing primary constraints (which role is played by relations \eqref{s13} here) to find the secondary constraints. The difference between these procedures lies in the fact that here we do not introduce the generalized momenta and consider the metric $g_{\m\n}$ as the external field, so its time derivatives are thus not related to dynamics.

If one takes the time derivative of \eqref{s13} and uses the dynamical equations \eqref{s14} together with \eqref{s13}, only one new equation appears, which, according to our terminology, is a constraint:
\disn{s14a}{
	e_{0 a} D_i e_k^a=0
	\nom}
(note that it is a part of \eqref{s10}).
The other appearing equations follow from \eqref{s13}, \eqref{s14} and  \eqref{s14a} at a given moment of time (i.e. without the usage of time derivative). Taking time derivative of \eqref{s14a}, in turn, leads to one more new constraint
\disn{s17}{
	\ta^{lm}=-\al^{lm}_a \biggl( \ta^{00}\al^{ika}\Bigl( (D_i e_{0b})(D_k e_0^b)+R_{0i0k}\Bigr) +2\ta^{0n}D_n e_0^a\biggr),
	\nom}
where $R_{\m\n\al\be}$ is a curvature tensor and the quantity denoted as $\al^{lm}_a$ is, in general case, uniquely defined by relations
\disn{s16}{
	\al^{ik}_a=\al^{ki}_a,\qquad
	\al^{ik}_a e^a_\m=0,\qquad
	\al^{ik}_a D_l e^a_m=\frac{1}{2}\ls \de^i_l\de^k_m+\de^i_m\de^k_l\rs.
	\nom}
If we use the constraint \eqref{s17} to exclude $\ta^{lm}$ from the set of the independent variables
(substituting it into equations \eqref{s14}), then it becomes not necessary to consider its time derivative.

As a result, we obtain that the remaining independent variables describing embedding matter are $e_\m^a$ and $\ta^{0\n}$, whose dynamics are governed by equations
\disn{s18.1}{
D_0 e_k^a=D_k e_0^a,
\nom}\vskip -2em
\disn{s18.2}{
D_0 e_0^a=\al^{ika}\Bigl( (D_i e_{0b})(D_k e_0^b)+R_{0i0k}\Bigr),
\nom}\vskip -2em
\disn{s18.3}{
D_0 \ta^{00}=-D_i \ta^{0i},
\nom}\vskip -2em
\disn{s18.4}{
D_0 \ta^{0m}=D_l \Biggl(\al^{lm}_a \biggl( \ta^{00}\al^{ika}\Bigl( (D_i e_{0b})(D_k e_0^b)+R_{0i0k}\Bigr) +2\ta^{0k}D_k e_0^a\biggr)\Biggr).
\nom}
The full set of constraints which are imposed on the independent variables at the initial moment of time, has the form
\disn{s19.1}{
e_\m^a e_{\n a}=g_{\m\n},
\nom}\vskip -2em
\disn{s19.2}{
D_i e_k^a=D_k e_i^a,\qquad
e_{0 a} D_i e_k^a=0.
\nom}
At other moments of time they become satisfied automatically.

\section{Nonrelativistic limit}\label{nonrel}
Let us analyze the equations of motion of embedding gravity in the \emph{non-relativistic in the bulk} limit \eqref{s7}.
We also assume that $T^{\m\n}$ and $j^\m_a$ (and therefore $\ta^{\m\n}$) are not so large,
so due to the small value of gravitational constant $\ka$ the r.h.s. of Einstein equations in \eqref{r3}
is also small, so the gravitational field turns out to be weak:
\disn{s20}{
	g_{\m\n}=\eta_{\m\n}+h_{\m\n},\qquad
	h_{\m\n}\to0,
	\nom}
where $\eta_{\m\n}$ is the Minkowski space metric.

Let us first find the solution in the limiting case when $\de j^\m_a=0$ (see \eqref{s7}) and $h_{\m\n}=0$, the corresponding quantities will be marked by a bar.
Since $\bar j^\m_I=0$ (here and hereafter $I,K,\ldots=1,\ldots,9$), we have $\bar\ta^{\m\n}\bar e_\n^I=0$.
The rank of the matrix $\bar e_\n^I$ can be either 4 or 3, since the rank of $\bar e_\n^a$ must be equal to 4.
In the first case $\ta^{\m\n}=0$, i.e. the embedding matter is absent.
Let us consider a second case, when a solution of the form
\disn{s25}{
	\bar e^a_0=\de^a_\np,\qquad
	\bar e^\np_k=0,\qquad
	\bar\ta^{\m\n}=\bar\rho_\ta \de^\m_0 \de^\n_0
	\nom}
is possible\footnote{It can be
show \cite{statja68}
that any solution in the second case can be brought to this form by the choice of coordinates.},
which corresponds to the static dust embedding matter.

Taking \eqref{s25} into account, we obtain the following solution from the constraints \eqref{s19.1},\eqref{s19.2}:
\disn{s21.4}{
	\bar e^I_k=\dd_k \bar y^I,
	\nom}
where $\bar y^I(x^i)$ is an embedding of a flat 3-dimensional metric in a nine-dimensional Euclidean space. Since 3-dimensional metric has 6 independent components, such embedding is paramet\-rized by three arbitrary functions.
In the generic case (which, for example, is realized when the initial values for the solution of dynamical equations \eqref{s18.1}-\eqref{s18.4}, which is restricted by the constraints \eqref{s19.1},\eqref{s19.2}, are not fine-tuned) this embedding has such form that there is a quantity $\bar\al^{ik}_a$ defined by \eqref{s16} (it can be noticed that $\bar\al^{ik}_\np=0$). This structure of embedding corresponds to the notion of the \emph{free embedding} introduced in the work \cite{bustamante}, if we consider it from the 3-dimensional point of view, and to the \emph{spatially free embedding} from the 4-dimensional point of view. The dynamical equations \eqref{s18.1}-\eqref{s18.4} in the considered limiting case are reduced to
\disn{s26}{
	\dd_0\bar e_k^I=0,\qquad
	\dd_0\bar\rho_\ta=0.
	\nom}
Geometrically, this case corresponds to the embedding function of the 4-dimensional Min\-kow\-ski metric in a 10-dimensional space: $\bar y^a(x^\m)=\{x^0,\bar y^I(x^i)\}$.

Now let us find small departures from the obtained limiting case, which correspond to the small but nonzero $\de j^\m_a$ and $h_{\m\n}$. Note that according to Einstein equations $h_{\m\n}\sim \ka$. We will denote the perturbation of any quantity, say, $F$, as  $\de F$, meaning by it the difference $F-\bar F$ between the quantity $F$ and its value $\bar{F}$ in the above limit. Considering all $\de F$ as small, we will neglect $(\de F)^2$ in comparison with $\de F$. Solving the constraints \eqref{s19.1},\eqref{s19.2} up to that order, we obtain that
\disn{s34}{
	\de e^\np_i=\dd_i w,\qquad
	\de e^I_i=\frac{1}{2}\dd_i\ls
	\bar\al^{ikI}\Bigl( (\dd_i w)(\dd_k w)-h_{ik}\Bigr)\rs,\no
	\de e_0^\np=\frac{1}{2}\ls h_{00}-\de e_0^I \de e_{0I}\rs,\qquad
	\de e_0^I=\bar e^{kI}\ls h_{0k}-\dd_k w\rs+\bar\al^{ikI}\ls\Ga_{ik}^0-\dd_i \dd_k w\rs,
	\nom}
where $\Ga^\al_{\m\n}$ are Christoffel symbols and $w$ is a small arbitrary function.
It should be noted that the solution of the constraint equations \eqref{s19.1},\eqref{s19.2} are, as a whole, parametrized by 4 arbitrary functions corresponding to a 3-dimensional surface embedded in 10-dimensional space. However, when we study the perturbations on this background, three of them are reduced to the modification of the three arbitrary functions mentioned above that describe the limiting case. Therefore it is sufficient to introduce only one function $w$ in the study of the perturbations.
Besides that, at the initial moment of time we have to arbitrarily define four unconstrained small perturbations $\de\ta^{0\n}$, but $\de\ta^{00}$ is reduced to the alteration of $\bar\rho_\ta$, so it is sufficient to consider only the perturbations  $\de\ta^{0k}$.
The components of perturbations of currents $\de j^\m_a$ have the form
\disn{s41}{
	\de j^{\m I}=
	\bar\rho_\ta \de^\m_0  \de e_0^I+\de \ta^{\m k}\bar e^I_k+\de \ta^{\m\n}\de e_\n^I
	\nom}
and as long as they remain small, the currents $j^\m_a$ are non-relativistic in the bulk.

Let us then analyze the dynamical equations \eqref{s18.1}-\eqref{s18.4} in the presence of the small perturbations from the limiting case. We assume that the gravitational field changes slow enough, i.e. $\dd_0 g_{\m\n}=o(\ka)$ (remind that $h_{\m\n}\sim \ka$). Later we will see that this assumption is consistent with the slow evolution of the r.h.s. of the Einstein equations. We will not consider the equations \eqref{s18.1},\eqref{s18.2} at $a=\np$, because the components $e_\m^\np$ at any moment can be expressed through other quantities from the diagonal part of the constraint \eqref{s19.1}, so it is not necessary to analyze their dynamics separately. As to the remaining equations \eqref{s18.1}-\eqref{s18.4} which have to be studied, we note that if the perturbations are chosen to be small enough at the initial time, namely, $w=o(\sqrt{\ka})$, $\ta^{0i}=o(\sqrt{\ka})$, then, taking \eqref{s34} into account, we can see that for some time the r.h.s. of the considered dynamical equations depend on time trivially (i.e., do not change or change linearly) at the leading order. Therefore their solutions also have to be trivially dependent on time (linearly or quadratically). However, after some time, namely, when $w\sim\ta^{0i}\sim\sqrt{\ka}$ (it will happen in $\Delta x^0\sim\ka^{-1/2}$) these r.h.s. will start to change nontrivially, and the dynamics will become nontrivial. To analyze this dynamics (which can be recognized as standard non-relativistic one), we need to change the time variable $x^0=ct$, where $c\sim \ka^{-1/2}$, so the usual speed of light can be taken as constant $c$, since $\ka=8\pi G/c^2$.

After this substitution the considered dynamical equations can be rewritten (up to corrections of order $1/c$) in the following form:
\disn{s55}{
	\dd_t \bar e_k^I=\dd_k \ga^I,\qquad
	\dd_t \ga^I=
	-\bar e^{kI}\dd_k\ff+\bar \al^{ikI}\Bigl( (\dd_i \ga_L)(\dd_k \ga^L)-\dd_i\dd_k\ff\Bigr),\qquad
	\dd_t \bar\rho_\ta=
	-\dd_i \ls\bar\rho_\ta v^i_\ta\rs,\no
	\dd_t \ls\bar\rho_\ta v^m_\ta\rs=-\bar\rho_\ta\dd_m\ff+
	\dd_l \Biggl(\bar\rho_\ta\bar \al^{lm}_L \biggl( \bar \al^{ikL}\Bigl( (\dd_i \ga_I)(\dd_k \ga^I)-\dd_i\dd_k\ff\Bigr) +2v^i_\ta\dd_i \ga^L\biggr)\Biggr),
	\nom}
where $\dd_t\equiv\dd/\dd t=c\dd_0$,
$\ga^I=c e_0^I$, $v_\ta^i=c\ta^{0i}/\ta^{00}$.
The field $\ff$ is the Newtonian gravitational potential corresponding to the distribution of matter with the density $\rho+\bar\rho_\ta$; it naturally appears in the solution of Einstein equation \eqref{r3} in the assumption that the ordinary matter is pressureless and non-relativistic, i.e. $T^{\m\n}=\rho \de^\m_0 \de^\n_0$ in the leading order of $1/c$.

Now let us take into account the solution of the constraints \eqref{s21.4} and \eqref{s34}, so we can express the obtained dynamical variables in the more compact form, using $\bar y^I$ and $\ps=cw$. After some calculation we obtain (again, up to $1/c$) \emph{the non-relativistic equations of motion} of embedding matter:
\disn{s66.1}{
\dd_t\bar{y}^I= \ga^I,\qquad
\dd_t\ps=\ff+\frac{1}{2}\ga^I\ga^I,
\nom}\vskip -2em
\disn{s66.2}{
\dd_t\bar\rho_\ta=
-\dd_i \ls\bar\rho_\ta v^i_\ta\rs,
\nom}\vskip -2em
\disn{s66.3}{
\bar\rho_\ta\ls\dd_t+v^i_\ta\dd_i\rs v^m_\ta=-\bar\rho_\ta\dd_m\ff+\ns+
\dd_l \Biggl(\bar\rho_\ta\biggl[v^l_\ta v^m_\ta+\bar \al^{lm}_L \biggl( \bar \al^{ik}_L\Bigl( (\dd_i \ga^I)(\dd_k \ga^I)+\dd_i\dd_k\ff\Bigr) +2v^i_\ta\dd_i \ga^L\biggr)\biggr]\Biggr),
\nom}
where
\disn{s67}{
	\ga^I=(\dd_k \bar y^I)\dd_k \ps+\bar\al^{ik}_I\dd_i \dd_k \ps,\qquad
	\bar\al^{ik}_I \dd_m \bar y^I=0,\qquad
	\bar\al^{ik}_I \dd_l\dd_m \bar y^I=\frac{1}{2}\ls \de^i_l\de^k_m+\de^i_m\de^k_l\rs,
	\nom}
and $\bar y^I$ is an embedding of a flat 3-dimensional metric (it can be checked that it remains flat at any time) in a 9-dimensional space. This embedding is parametrized by three independent functions, whose explicit form we do not succeed to obtain. These three functions, alongside with $\ps$, $\bar\rho_\ta$ and $v^i_\ta$, are the independent variables that describe the embedding matter, and do not affected by any constraints anymore. The equation \eqref{s66.2} is the usual conservation law of this matter, whereas \eqref{s66.3} defines its law of motion. The r.h.s. of it is a total force (in a unit volume) acting on a "particles" of matter. Its first term is an ordinary gravitational force corresponding to the Newtonian approximation, whereas the remaining ones can be interpreted as some self-interaction force of the embedding matter that depends not only on its usual characteristics (density $\bar\rho_\ta$ and velocity $v^i_\ta$), but also on the additional ones  ($\ps$ and $\bar y^I$), whose dynamics are governed by the equations \eqref{s66.1}.

It is interesting from the geometrical point of view to write down the embedding function of 4-dimensional surface $y^a(x^\m)$ corresponding to the obtained non-relativistic regime:
\disn{s68}{
	y^0=x^0+\frac{1}{c}\,\ps\ls\frac{x^0}{c},x^i\rs+o\ls\frac{1}{c^2}\rs,\no
	y^I=\bar y^I\ls\frac{x^0}{c},x^i\rs+\frac{1}{c^2}\bar\al^{lmI}\ls\frac{1}{2}(\dd_l \ps)(\dd_m \ps)-\ff\de_{lm}\rs+o\ls\frac{1}{c^2}\rs.
	\nom}
One can easily check that the corresponding induced metric reproduces the standard non-relativistic expression
\disn{s68a}{
g_{\m\n}=\eta_{\m\n}+\frac{2\ff}{c^2}\de_{\m\n}+o\ls\frac{1}{c^2}\rs.
	\nom}

\section{Discussion}\label{zakl}
We study the possibility to explain the mystery of DM through the transition from GR to \emph{embedding gravity}, which is a modified gravity based on a simple geometric principle similar to the one that is used in the string theory. Since the discrepancies between the observations and the GR are explained quite well by the hypothesis that DM is cold and dust-like,
we reformulate the equations of motion of embedding gravity as Einstein equations with an additional contribution
of embedding matter and search for a class of solution, for which this matter turns out to be non-relativistic.
The non-relativistic regime of motion of the fictitious embedding matter turns out to be stable in the following sense. If the initial values are chosen according to the non-relativistic case, the resulting motion remains non-relativistic during the dynamics. Especially remarkable is the fact that in the non-relativistic limit, embedding gravity possesses some self-interaction besides the ordinary interaction with gravity, as in the SIDM models mentioned in the Introduction.
It could be useful in the context of solving the core-cusp problem
and similar issues existing in the $\La$CDM model
on the galactic scale,
such as the missing satellites and too-big-to-fail problems.

As noted in the introduction, the approach to solving the DM mystery considered in this work is close to the mimetic gravity approach.
It should be emphasized that within the framework of mimetic gravity, progress in obtaining a correct description of DM occurs after appropriate modification of the action (see for example the recent works \cite{1510.02284,1803.02620}) since the original formulation of mimetic gravity is not complex enough. In contrast to this, embedding gravity is initially a very complex theory, which can make it possible to do without its additional complications.
But, on the other hand, the complexity of the theory does not allow us to quickly answer many of the questions that arise.
For example, it is difficult to say for sure whether the additional degrees of freedom of embedding gravity corresponding to embedding matter are ghosts.
From the fact that in the considered nonrelativistic limit they are reduced to density $\bar\rho_\ta$ and velocity $v^i_\ta$ it can be assumed that this is not the case, but this issue requires additional careful study.

When the embedding gravity is written as GR with additional degrees of freedom, the embedding matter is effectively parametrized by eight fields: besides its density $\bar\rho_\ta$ and velocity $v^i_\ta$, the field $\ps$ takes part in the description, as well as embedding function  $\bar y^I$ of a flat 3-dimensional metric in a 9-dimensional space, which definition is equivalent to the choice of three functions. These four additional degrees of freedom have their own dynamics and affect the self-interaction of embedding matter.

The behavior of embedding matter is determined by initial values of the non-relativistic equations \eqref{s66.1}-\eqref{s66.3}, which have to be imposed at the beginning of the non-relativistic regime. These values themselves must be defined by the dynamics of the previous (relativistic) regime, which has to be studied separately.
Only after such a study will it be possible to answer any questions regarding the behavior of embedding matter.
In particular, it will be possible to derive an effective self-interaction cross-section arising for an embedding matter in the nonrelativistic limit, since in the case under consideration it is not possible to evaluate it directly from the action.
Especially interesting is the question of the description of inflation, in particular, whether it is possible to obtain the inflaton using the additional degrees of freedom of the embedding gravity? If so, there is no need to add the inflaton in theory as a separate entity.

It should be noted that the choice of the embedding function of the 3-space in this work is drastically different from the one that was used at cosmological scale in the works \cite{davids01,statja26}, where it has the minimal codimension (i.e., a 3-surface was embedded in a flat 4-bulk). In this work, we assume that, for the generic initial values, this 3-surface turns out to be "unfolded" in 9-dimensional flat bulk, which corresponds to the existence of the quantity $\bar \al^{ik}_I$ defined by \eqref{s67}. The choice of such surface, which would be "unfolded"{}, but, as a result of inflation, would correspond to the observed homogeneity and isotropy of the 3-dimensional space, could be the subject of future study.

{\bf Acknowledgements.}
The author is grateful to A.~Golovnev and A.~Sheykin for useful discussion.
The work is supported by RFBR Grant No.~20-01-00081.


\end{document}